\documentstyle[floats,aps,eqsecnum]{revtex}
\renewcommand{\today}{Nov. 18, 1994}
\newcommand{\Sp}{\,}
\newcommand{\intl}{\int\limits}
\newcommand{\svph}{\vphantom{\sum_N}}
\newcommand{\bvph}{\vphantom{\sum_N^N}}

\begin{document}
\draft
\catcode`\@=11
\def\preprint#1{\def\@preprint{\noindent\hfill\hbox{#1}\vskip 10pt}}
\catcode`\@=12
\preprint{\vbox{\hbox{cond-mat/9411078}\hbox{Subm. to Nucl.\ Phys.\ B}}}
\twocolumn[\hsize\textwidth\columnwidth\hsize\csname%
@twocolumnfalse\endcsname
\title{Fractional Spin for Quantum Hall Effect Quasiparticles}
\author{T. Einarsson}
\address{Laboratoire de Physique des Solides,
Universit\'e Paris-Sud, F-91405 Orsay, France}
\author{S.~L. Sondhi}
\address{Department of Physics, University of Illinois at
Urbana-Champaign, Urbana, IL 61801-3080, USA}
\author{S.~M. Girvin}
\address{Department of Physics, Indiana
University, Bloomington, IN 47405, USA}
\author{D.~P. Arovas}
\address{Department of Physics, University of California at
San Diego, La Jolla, CA 92093, USA}
\date{\today}
\maketitle
\begin{abstract}
We investigate the issue of whether quasiparticles in the fractional
quantum Hall effect possess a fractional intrinsic spin. The presence
of such a spin $S$ is suggested by the spin-statistics relation
$S=\theta/2\pi$, with $\theta$ being the statistical angle, and, on a
sphere, is required for consistent quantization of one or more
quasiparticles.  By performing Berry-phase calculations for
quasiparticles on a sphere we find that there are two terms, of
different origin, that couple to the curvature and can be interpreted
as parts of the quasiparticle spin.  One, due to {\em
self-interaction}, has the same value for both the quasihole and
quasielectron, and fulfills the spin-statistics relation.  The other
is a kinematical effect and has opposite signs for the quasihole and
quasielectron.  The total spin thus agrees with a generalized
spin-statistics theorem $\textstyle{\frac{1}{2}}(S_{\rm qh} + S_{\rm
qe}) = \theta/2\pi$.  On the plane, we do not find any corresponding
terms.
\end{abstract}
\vspace*{10mm}]

\section{Introduction}
The fractional quantum Hall effect (FQHE) reflects the existence of a
family of strongly correlated states of two-dimensional electrons in a
transverse magnetic field. Perhaps the most striking consequence of
the strong correlation in these states is that they support
quasiparticle excitations with fractional quantum numbers. Their
fractional charge was discovered by Laughlin in his initial work on
the problem~\cite{Laugh1} and subsequently Halperin~\cite{Halp1}, in
his construction of the hierarchy, pointed out that they are most
naturally assigned fractional statistics. These results were
subsequently confirmed by Arovas, Schrieffer, and Wilczek~\cite{Arov1}
by a direct Berry-phase analysis.

In this communication, we are concerned with a third fractional
quantum number for FQHE quasiparticles --- a fractional intrinsic
angular momentum or spin. Two remarks are immediately in order. First,
we emphasize that the spin of the quasiparticles has, logically,
nothing to do with the intrinsic spin-$\textstyle{\frac{1}{2}}$ of the
electron itself; it arises even for spinless electrons. (Consequently,
in most of the subsequent analysis we restrict ourselves to the
spinless problem and only take account of the, benign, complications
of the true electron spins at the end.) Second, this fractional spin
is evidently not an ordinary ${\rm SU(2)}$ spin, but rather an ${\rm
SO(2)}$ angular momentum perpendicular to the two-dimensional surface
the electrons inhabit.

There are two distinct reasons why one might expect to find such a
fractional spin for the quasiparticles. The first has to do with a
general expectation of a spin-statistics relation for
fractional-statistics particles and goes back to Wilczek's first
papers about anyons, which he called fractional-spin
particles~\cite{Wilczek}.  In model calculations, the spin arises from
the self-interaction of the anyon's own charge and flux-tube and,
while its value $S$ is not quantized, it fulfills the spin-statistics
relation $S=\theta/2\pi$ for anyons with exchange statistics
$\theta$. For general non-relativistic systems there is no proof of
this relation, particularly in a background magnetic field which might
well introduce extra phases. There is, however, a fairly compelling
argument for a generalized spin-statistics relation, which we now
sketch.

Consider a cluster of one anyon $a$ and one antianyon $\bar{a}$.
Since $a$ and $\bar{a}$ have opposite charge and opposite flux, their
total flux and charge is zero. In the standard case, all the quantum
numbers of this cluster are zero.  The Berry phase for the rotation of
this cluster by an angle of $2\pi$ around its center of mass is then
unity, but, in the spirit of Thouless and Wu~\cite{Thou1}, it can also
be calculated as $\exp[i(- 2\theta + 2\pi S_a+2\pi S_{\bar{a}})] = 1$,
where $-2\theta$ is the phase for taking the anyon around the
antianyon (remember that the antianyon has opposite ``charge'' and
``flux'' with respect to the anyon) and $S_a$ ($S_{\bar{a}}$) is the
anyon (antianyon) spin.  For this to be valid, it is necessary that
the part of the spin that is even under charge conjugation satisfies
\begin{equation} \label{eq:gen-sp-stat}
S_{\rm even}\equiv \textstyle{\frac{1}{2}}(S_a + S_{\bar{a}})
=\frac{\theta}{2\pi} \pmod{\textstyle{\frac{1}{2}}} \Sp,
\end{equation}
while the odd part, $S_{\rm odd} = \textstyle{\frac{1}{2}}(S_a -
S_{\bar{a}})$, is unconstrained.\footnote{The essence of this idea was
first put forward by A. Goldhaber (unpublished).  T. H. Hansson and
A. Karlhede have also noted the arbitrariness of $S_{\rm odd}$
(unpublished).}  This generalized spin-statistics relation should be
valid in all systems with anyons and antianyons, but it is not clear
that it is applicable to the FQHE case.  Indeed, if we consider a
quasihole-quasielectron pair on a plane, and rotate the relative
coordinate $(z_a-z_{\bar{a}})$ by $2\pi$, there is no spin
contribution at all; the rotation of the cluster is not a rigid
operation.  One could instead try to rotate the whole system, but here
the calculation is plagued by a divergent angular
momentum~\cite{Sond1}, and there is not, to the best of our knowledge,
any convincing argument yielding a spin in agreement with
Eq.~(\ref{eq:gen-sp-stat}).

The second reason for an intrinsic spin involves consistency
conditions on a system of fractionally charged anyons on curved and
closed surfaces~\cite{EinaLi}.  These conditions typically occur
because, on a closed surface, one can view either the enclosed region
to the left {\em or\/} to the right as the interior of a loop, and the
corresponding geometric phases must agree up to a multiple of
$2\pi$. In general, this leads to relations between the statistics
$\theta$, the spin $S$, the fractional charge $e^{*}$, the total
magnetic flux through the surface $N_\phi$, and the total curvature of
the surface.  These types of arguments do apply to FQH systems on
curved and closed surfaces.  The study of such systems, begun by
Haldane in his seminal study of the FQHE on a sphere~\cite{Hald3}, has
been argued by Wen~\cite{Wen9} to be a fundamental way of
characterizing ``topological'' fluids of which the Hall fluids are an
example.

For our purposes, the interesting questions have to do with the
subtleties of quantizing the effective action for a collection of
fractionally charged anyons~\cite{Bala8} such as the quasiparticles
that are present away from the commensurate fillings.  As an example,
it cannot be true that the total Berry phase for one quasihole being
transported around a loop on a sphere is simply the Aharonov-Bohm (AB)
phase as it is on the plane.  This is because the consistency
relations demand that $N_\phi e^{*}$ is an integer, which is not the
case (see Sec.~\ref{sec:AB}).  In an effective theory describing the
quasiholes, there must thus be another geometric phase which together
with the AB-phase gives a well-defined overall phase. Such a phase
should come out of the calculation of the total Berry phase, which is
always well-defined.

In this paper we perform a detailed calculation of the Berry phase for
a closed loop for a single quasiparticle on a sphere.  For the
Laughlin fractions, with filling $\nu=1/m$ ($m$ odd integer), we find
that the additional phase can be interpreted as coming from a spin of
the quasielectron/quasihole with the value $S=\pm
\textstyle{\frac{1}{2}} + 1/2m$.  What one sees from our derivation is
that the second, charge-conjugation-even, part is due to
self-interaction, while the odd part is a kinematical effect due
(heuristically) to the coupling of the cyclotron orbits to the
curvature~\cite{Wen20}.  Since the self-interaction part fulfills the
standard spin-statistics relation $S=\theta/2\pi$, the total spin
fulfills relation (\ref{eq:gen-sp-stat}). This is true also for the
hierarchy states. Thus, we find that the general spin-statistics
relation (\ref{eq:gen-sp-stat}) is valid for FQH quasiparticles on a
sphere, even though the cluster-argument for this relation does not
apply.  As discussed above, on the plane there is no spin or
self-interaction contribution.

We emphasize that the above two motivations for a spin are logically
independent, at least for non-relativistic systems.  This becomes
clear in a path-integral description, where the action can include
terms that measure the self-linking of the world lines of the
particles in (2+1)-dimensional spacetime (corresponding to the first
motivation) and terms that measure the curvature enclosed by their
projection onto the 2-dimensional space (corresponding to the second
motivation).%
\footnote{The notion of self-linking is somewhat ambiguous in the case
of non-relativistic anyon theories. For a discussion of the
path-integral formulation of the relativistic case, see, e.g.,
Ref.~\protect{\cite{Hanss}}.}

Our principal conclusions are 1) for quantum Hall systems there are no
terms in the quasihole/electron actions proportional to self-linking
2) there {\em are\/} terms proportional to the curvature of the
surface they inhabit, and 3) the coefficients of these terms, the
spins of the quasiparticles, satisfy the generalized spin-statistics
relation (\ref{eq:gen-sp-stat}).

There has been previous work on this issue. Sondhi and
Kivelson~\cite{Sond1} considered planar FQHE systems and investigated
spins defined as integrals of angular momentum operators restricted to
the region occupied by an isolated quasiparticle at rest but,
correctly, failed to find one that satisfied that spin-statistics
relation. Our current understanding is that given the lack of
dynamical significance for a spin in planar systems (the absence of
self-linking) there is no reason to expect a localized fractional
angular momentum consistent with the statistics. Einarsson and
Girvin~\cite{Torbjorn2}, considered topological definitions more in
the spirit of this paper (albeit on the plane) which is, in a sense,
the completion of their program.  In their work on effective theories
of the FQHE on curved surfaces~\cite{Wen20}, Wen and Zee focussed on
the anti-symmetric part of the quasiparticle spin (their orbital spin)
though they also noted that an intrinsic spin ought to be present on
general grounds.  Li derived the existence of the spin term by
demanding that the wavefunctions of hierarchical daughter states be
rotationally symmetric~\cite{Li3} on the sphere.  A treatment of the
spin terms in the effective theories was then presented by Lee and
Wen~\cite{Lee9} in the bosonic Chern-Simons approach, but as they
inferred the self-interaction piece directly from the Chern-Simons
action, which has ultraviolet ambiguities~\cite{Polya-witten}, their
derivation cannot be considered microscopic.

The paper is organized as follows. We begin by reviewing the formalism
for the sphere and the general constraint on the values of the spin
that it imposes. Next we describe the Berry-phase calculation on the
sphere using the explicit wavefunction. We also explain why a
previously reported attempt at this calculation by Li~\cite{Li3} is
incorrect. We then reproduce our results by a different route which
generalizes readily to the hierarchy. We conclude with a summary and
some remarks on the inclusion of the intrinsic spin of the electrons.

\section{FQHE on the Sphere}

\subsection{Formalism}

We briefly review the formalism for the QHE on the sphere due to
Haldane~\cite{Hald3,Fano1}, largely because we use a different set of
conventions.  We consider a system of $N_e$ electrons on a sphere of
radius $R=1$ that encloses a magnetic monopole. The strength of the
monopole is quantized according to the Dirac condition which requires
that the integrated flux across the sphere, which we take to be
positive, $N_{\phi} \equiv 2 q = 4\pi B$ be an integer number of flux
quanta $\Phi_0 = hc/e \equiv 1$.  The single-particle states form
representations of the rotation group $SU(2)$ with $l = q, q+1
\ldots$, where each value of $l$ occurs precisely once and corresponds
to one particular Landau level.  The lowest Landau level (LLL) has
$l=q$ and contains $2q+1 = N_\phi+1$ degenerate states, one more than
for the same flux on the plane or torus.

Explicit wavefunctions require a choice of gauge for the vector
potential of the monopole and hence a choice of its Dirac strings. In
the QHE literature it is conventional to follow Haldane in choosing
\begin{equation} \label{eq:haldanegauge}
{\bf A} = - B \frac{\cos \theta}{r \sin \theta} \hat{\varphi} \Sp,
\end{equation}
which involves a string through each pole of the sphere. This is
inconvenient for our purposes for it produces spurious additions to
the Berry phase compared to the planar case. These spurious
contributions arise because even the stereographically projected
planar vector potential has a singularity at the origin. We therefore
revert to a more traditional gauge choice, the ``Dirac'' gauge,
\begin{equation} \label{eq:diracgauge}
{\bf A} = B \frac{1 - \cos\theta}{r \sin \theta} \hat{\varphi} \Sp,
\end{equation}
with a single string through the south pole.

We also depart from the standard conventions by requiring that the LLL
consist of holomorphic functions. As on the plane, this is
accomplished either by reversing the sign of the magnetic field in
Eq.~(\ref{eq:diracgauge}) or equivalently, as in this paper, by
considering a system of charge $+e$ electrons
(positrons).\footnote{Since we already need a monopole this does not
greatly add to the difficulty of realizing the system experimentally.}
Then the LLL wavefunctions are conveniently written using the
spinorial coordinates $u= \cos(\theta/2)$, $v=\sin(\theta/2) e^{i
\varphi}$ as
\begin{equation}
\psi_{q m} = \left[ \frac{2 q +1}{4 \pi} {2 q \choose q + m}
\right]^{1/2} u^{q-m} v^{q+m} \Sp,
\end{equation}
where $m$ is the eigenvalue of the z-component of the angular momentum
operators,
\begin{mathletters}
\begin{eqnarray}
-L_x &=& \frac{1}{2} \left( v \partial_u + u \partial_v \right)\Sp,\\
L_y &=& \frac{i}{2} \left( v \partial_u - u \partial_v \right)\Sp,\\
L_z &=& \frac{1}{2} \left( v \partial_v - u \partial_u \right) \Sp.
\end{eqnarray}
\end{mathletters}
Stereographic projection to the plane from the north pole is
accomplished by the map $(u,v) \rightarrow z=v/u=
\tan(\theta/2)\exp(i\varphi)$.

The rotationally invariant ($L=0$) many-particle wavefunctions,
\begin{equation} \label{eq:L-sphere}
| m \rangle = \prod_{i<j}(u_i v_j -u_j v_i)^m \Sp,
\end{equation}
describe the ground state when $N_\phi = m(N_e-1)$, and are the
generalizations to the sphere of the Laughlin wavefunctions for the
fractions $\nu = 1/m$. The shift in the relation between $N_e$ and
$N_\phi$ with respect to the planar problem is important in what
follows since it gives rise to the odd component of the spin.
Finally, the one-quasihole state with the quasihole center given by
the spinor coordinates $(\alpha,\beta)$ is described by
\begin{equation} \label{eq:qh-sphere}
| \alpha,\beta \rangle = \prod_{j=1}^{N_e} (\beta u_j - \alpha v_j) |
m \rangle \Sp,
\end{equation}
where $N_\phi = m(N_e-1)+1$, i.e., it is created by increasing the
total flux by one quantum. These states have $L=N_e/2$, consistent
with their fractional charge~\cite{Hald3}, and the angular momentum is
{\em maximally\/} polarized along the outward-pointing axis through
($\alpha, \beta$).  More generally, the flux-number relation for a
system with $N_{\rm qh}$ quasiholes and $N_{\rm qe}$ quasielectrons is
\begin{equation}
N_{\phi} = m(N_e-1) + N_{\rm qh} - N_{\rm
qe}\Sp.\label{eq:flux-relation}
\end{equation}

\subsection{Aharonov-Bohm phase for a Quasiparticle} \label{sec:AB}
We can now precisely state the problem of consistently quantizing
quasiparticle actions on the sphere.  When geometric phases are
considered on a sphere, there is an ambiguity in what is considered as
being inside and outside a closed loop. More specifically, one may
calculate two geometric phases $\gamma_{}^{{\rm I}}$ and
$\gamma_{}^{{\rm II}}$ by considering the region to the left and the
right (with opposite orientation of the curve and thereby an
additional sign) as the interior of the loop. These phases must agree
up to a multiple of $2\pi$, i.e.,
\begin{equation}\label{eq:invariance}
\gamma(\Omega) = -\gamma(4\pi-\Omega) \pmod{2\pi}\Sp.
\end{equation}

Suppose we have integrated out all the electrons so that we have an
effective theory of one quasihole in a magnetic field.  We now
consider the Aharonov-Bohm phases for one charge-$e^{*}$ object being
transported around a loop enclosing the solid angle $\Omega$ to the
left, and $4\pi - \Omega$ to the right.  These are
\begin{mathletters}
\begin{eqnarray}
\gamma_{{\rm AB}}^{}(\Omega) &=& 2\pi
	\frac{e^{*}}{e}\frac{\Omega}{4\pi} N_{\phi} \label{eq:AB-I}
	\Sp,\\ -\gamma_{{\rm AB}}^{}(4\pi-\Omega) &=& \gamma_{{\rm
	AB}}^{}(\Omega) - 2\pi \frac{e^{*}}{e}N_{\phi} \Sp,
\end{eqnarray}
\end{mathletters}
where $\frac{\Omega}{4\pi}N_\phi$ is the enclosed magnetic flux.
{}From the consistency relation (\ref{eq:invariance}), together with
the quasihole charge $e^{*} = -e/m$, it must be that $N_{\phi}$ is a
multiple of $m$.  This does not agree with
Eq.~(\ref{eq:flux-relation}) with $N_{\rm qh}=1$, $N_{\rm qe}=0$, for
the Haldane-Laughlin wavefunction. Thus, in contrast to the plane, the
actual total geometric phase of a one-quasihole system, must differ
{}from the AB phase for one fractionally charged particle. This
discrepancy originates in the sphere being curved and closed, as we
will see below.  What is the additional phase needed to make the total
geometric phase well defined? A simple calculation shows that the
phase
\begin{equation} \label{eq:cnst-spin}
	\gamma_{\rm Spin} = (\frac{1}{2m} + \frac{n}{2})\Omega \Sp,
\end{equation}
with $n$ being an arbitrary integer, does the job.\footnote{For
multi-valued wavefunctions one could have a more complicated
relation.}

The problem we have just discussed is a consequence of the fractional
charge of the quasihole.  If one instead studies the case of $m$
quasiholes, there is no problem with the fractional charge and the
total flux, but instead the braid relations are inconsistent unless an
additional phase is present. This phase should be proportional to the
total curvature $\Omega_{\rm tot} = 4\pi$, and one finds~\cite{EinaLi}
\begin{equation} \label{eq:cnst-spin2}
\gamma_{\rm Spin} = (\frac{1}{2m}+\frac{n}{2})\Omega_{\rm tot}\Sp,
\end{equation}
so that both the fractional charge and the fractional statistics need
the same spin-phase for consistency.

\section{The Berry phase on the sphere}
To see how the problem posed in the last section is resolved, we will
now calculate the Berry phase for transporting one quasihole around a
closed loop on the sphere in two different ways. The first is a
generalization of the Arovas-Schrieffer-Wilczek (ASW) method on the
plane~\cite{Arov1} to the sphere.  The second uses the mapping of the
whole system to a spin-problem and Berry's original calculation of the
phase for twisting a spin around a closed loop~\cite{Berr1}. The
answers are, of course, identical but the first has the virtue of
separating the background and the self-interaction contributions from
each other.  The latter has the advantage of computational simplicity
and confirms the former.

\subsection{ASW calculation}
We consider the transport of a quasihole at a fixed latitude on the
sphere,
$(\alpha,\beta) = (\cos\textstyle{\frac{\theta_0}{2}},
\sin\textstyle{\frac{\theta_0}{2}}e^{it})$
for $t:\, 0 \rightarrow 2\pi$.  The time derivative of the Berry phase
for the wavefunction (\ref{eq:qh-sphere}) is
\begin{eqnarray} \label{eq:dgam-dt}
\frac{d\gamma_{{\rm B}}^{}}{dt} &=& i\langle \alpha,\beta |
\frac{d(\alpha,\beta)}{dt} \rangle\nonumber\\
&=& i\langle \alpha,\beta |
\frac{d}{dt}\left\{\sum_{i=1}^{N_e} \ln(\beta u_i - \alpha
v_i)\right\} | \alpha,\beta \rangle \nonumber\\
&=& i\!\int\! d\Omega \,
\frac{d}{dt} \ln(\beta u - \alpha v) \, \langle \alpha,\beta |
\sum_{i=1}^{N_e} \delta({\hat{\bf\Omega}}-{\hat{\bf\Omega}}_i) |
\alpha,\beta \rangle\nonumber\\
&=& i\!\int d\Omega \, \frac{d}{dt} \ln(\beta u -
\alpha v)\, \langle\rho_{\alpha,\beta}(u,v)\rangle \Sp.
\end{eqnarray}
The expectation value of the electron density can be expressed as
$\langle\rho_{\alpha,\beta}(u,v)\rangle =\rho_0 +
\delta\rho_{\alpha,\beta} (u,v)$, where $\rho_0$ is the constant
density away from the quasihole and $\delta\rho_{\alpha,\beta}(u,v)$
is the density deviation in the vicinity of the quasihole itself.
Because the sphere is a closed surface, the expulsion of $1/m$
electrons {}from the quasihole area is reflected in an increase in the
average electron density, and thus~\cite{Hald9}
\begin{equation} \label{eq:rho-background}
\rho_0 = \frac{N_e + \frac{1}{m}}{4\pi} \Sp.
\end{equation}
Also, note that the normalization of $| \alpha,\beta \rangle$ is
independent of $(\alpha,\beta)$ for one quasihole, and is fixed by the
electron density.

Using the ``rotational'' symmetry in $t$, we can calculate the Berry
phase as
\begin{equation} \label{eq:gammaB}
\gamma_{{\rm B}}^{} = \int_0^{2\pi}\!\! dt \; \frac{d\gamma_{{\rm
B}}^{}}{dt} = 2 \pi \left. \frac{d\gamma_{{\rm
B}}^{}}{dt}\right|_{t=0} \Sp.
\end{equation}
The logarithmic derivative in spherical coordinates is
\FL
\begin{equation} \label{eq:dln-dt}
\frac{d}{dt} \ln(\beta u - \alpha v) = \frac{i
   \sin(\frac{\theta_0}{2})\cos(\frac{\theta}{2})e^{it} }
   {\sin(\frac{\theta_0}{2})\cos(\frac{\theta}{2})e^{it} -
   \cos(\frac{\theta_0}{2})\sin(\frac{\theta}{2})\,e^{i\varphi}}\Sp.
\end{equation}
Inserting its value at $t=0$ in (\ref{eq:dgam-dt}), we find that
\begin{eqnarray} \label{eq:dgam-int}
\left. \frac{d\gamma_{{\rm B}}^{}}{dt}\right|_{t=0} &=& -\int_0^\pi\!\!
d\theta\,\sin\theta \int_0^{2\pi}\!\! d\varphi \, \frac{1}{1 -
\cot(\frac{\theta_0}{2})\tan(\frac{\theta}{2})\,e^{i\varphi}}\nonumber\\
&&\times[\rho_0+\delta\rho_{\theta_0,t=0}(\theta,\varphi)] \Sp.
\end{eqnarray}

We first calculate the contribution from the background density
$\rho_0$.  Performing the $\varphi$-integral, we obtain
\begin{eqnarray} \label{eq:dgam0}
\left.\frac{d\gamma_{{\rm B}}^{0}}{dt}\right|_{t=0}&=&
 \pi\rho_0\int_0^{\pi}\!\!d\theta\, \sin(\theta)\,
[{\rm sign}(\theta-\theta_0)-1]
 \nonumber\\ &=& - \rho_0\,\Omega\,,
\end{eqnarray}
where $\Omega=2\pi[1-\cos(\theta_0)]$ is the solid angle enclosed (to
the left) by going around the latitude $\theta_0$.\footnote{Note that
we have integrated over the whole sphere, but that only the density to
the left contributes. This is not true in the Haldane gauge
(\ref{eq:haldanegauge}), but the total Berry phases (when including
the $\delta\rho$-term) are the same in both cases.}  As on the plane,
we therefore have
\begin{equation} \label{eq:encl-dens}
\gamma_{{\rm B}}^{0} = - 2\pi \langle n_e \rangle_\Omega \Sp,
\end{equation}
where $\langle n_e \rangle_\Omega$ denotes the total number of
enclosed electrons.\footnote{There is a sign error in Eq. (10) in the
ASW paper.}

Using the relation $N_e = (N_\phi-1)/m + 1$, we find that
\begin{equation} \label{eq:gam0}
\gamma_{{\rm B}}^{0} = -2\pi \frac{1}{m} \frac{\Omega}{4\pi} N_{\phi}
-\frac{\Omega}{2} \Sp,
\end{equation}
so that in addition to the Aharonov-Bohm part (\ref{eq:AB-I}), there
is a part proportional to the enclosed curvature $\Omega$.  The
curvature part can be traced back to the $-1$ offset in the
electron-flux relation $m(N_e-1)=N_\phi$. It is due to the coupling of
the electron cyclotron motion to the curvature and contributes
$\textstyle{\frac{1}{2}}$ to the total spin, consistent with relation
(\ref{eq:cnst-spin}).

Next we consider the contribution from $\delta\rho$. On a plane this
is known to vanish. This can be seen by repeating the calculation of
Arovas {\it et al.\/}\ for a circular charge distribution centered
around the quasihole coordinate $z_0$. Translating the center of the
coordinate system to $z_0$ and employing the circular symmetry gives
$d(\delta\gamma_{{\rm B}}^{})/dt=0$~\cite{Blok3}.  On the sphere, this
is not the case. In fact, the transport of the charge distribution
couples to the curvature and induces an effective rotation of the
quasihole around its center. This is a type of self-interaction where
the smeared-out electron density partly goes around the smeared-out
vorticity of the quasihole itself. To see how this works, we begin by
considering a simple form for the quasihole profile, namely, we assume
that $\delta\rho_{\theta_0,t=0}(\theta,\varphi)$ is constant inside a
circle of (geodesic) radius $\beta$. The area is then
$A(\beta)=2\pi(1-\cos\beta)$, and to make the total charge equal to
$-1/m$, the density deviation is $\delta\rho_0=-1/mA(\beta)$.  The
second term of Eq.~(\ref{eq:dgam-int}) then gives
\begin{equation} \label{eq:gamd}
\gamma_{{\rm B}}^{\delta} =
\frac{2\pi}{mA(\beta)}\intl_{\theta_0-\beta}^{\theta_0+\beta}\!\!
d\theta\, \sin(\theta)
\!\!\intl_{-\varphi_1(\theta)}^{\varphi_1(\theta)}\!\! d\varphi \,
\frac{1}{1 -
\cot\frac{\theta_0}{2}\tan\frac{\theta}{2}\,e^{i\varphi}}\Sp,
\end{equation}
where
\begin{equation} \label{eq:deltaphi}
\varphi_1(\theta) =
\arccos[\frac{\cos\beta-\cos\theta\cos\theta_0}{\sin\theta\sin\theta_0}]
\Sp.
\end{equation}
The inner integral can be performed analytically, but yields an
unwieldy integral in $\theta$ which we have solved analytically for
small $\beta$ (see the Appendix), and numerically for various values
of $\theta_0$ and $\beta$, with $\beta \leq \theta_0,\pi-\theta_0$.
The answer is in fact independent of $\beta$ and is given by
\begin{equation} \label{eq:gamS}
\gamma_{{\rm B}}^{\delta} = \frac{\Omega}{2 m}\Sp.
\end{equation}
Since the normalized contribution is independent of the radius
$\beta$, the density profile has no influence on the phase change, and
(\ref{eq:gamd}) is valid for {\em any\/} circularly symmetric
quasihole with charge $-1/m$.\footnote{In fact, the charge
distribution does not even have to be circularly symmetric.}  The
self-interaction part of the Berry phase is the sought
``spin-statistics'' piece of the spin phase (\ref{eq:cnst-spin}).

A heuristic argument suggests the origin of this term.  The basic idea
is that on a curved surface, the adiabatic transport of the
quasiparticle will effectively induce an adiabatic rotation of the
quasiparticle around its center. Assuming ordinary parallel transport
to be valid, the transport of the center around a loop enclosing the
solid angle (= curvature) $\Omega$ induces a rotation angle $\Omega$.
The geometric phase accumulated in this process can be used to define
a spin $S$, perpendicular to the surface, by Berry's
relation~\cite{Berr1}
\begin{equation} \label{eq:Berry-Berry}
\gamma(\Omega) = - S\, \Omega\Sp.
\end{equation}
Now consider a $2 \pi$ rotation of a quasiparticle.  As a
quasiparticle corresponds to a deviation in the charge distribution,
there will be a self-interaction as the quasihole is rotated around
its center. The outermost parts of the quasihole will move around the
total density-deviation, whereas the inner parts only move around a
small part of that deviation.  If we introduce $\rho_-$, a coordinate
denoting the total quasiparticle weight inside a certain radius, we
have from Eq.~(\ref{eq:encl-dens}) that the contribution from a thin
cylindrical slice $d\rho_-$ of the quasihole is
$-2\pi(-\rho_-)[d\rho_-/(\frac{1}{m})]$. The total phase is then
obtained as the integral
\begin{equation} \label{eq:spin-berry2}
   \gamma(2\pi) = 2\pi m \int_0^{1/m} \rho_- \, d\rho_- = \pi/m\Sp.
\end{equation}
For a general angle $\Omega$, we have $\gamma(\Omega) = \Omega/2m$ in
accordance with (\ref{eq:cnst-spin}), and from (\ref{eq:Berry-Berry})
we have that the self-interaction part of the spin is $S_{\rm s-i}
=-1/2m$.\footnote{We note that in this interpretation the dynamics on
the plane, unlike on the sphere, do not lead to rotations of the
quasiparticle.}

Collecting terms, we find that the total Berry phase is
\begin{eqnarray} \label{eq:tot-berry-qh}
 \gamma_{{\rm B}}^{\rm qh} &=& \gamma_{{\rm B}}^{0} +
 \gamma_{{\rm B}}^{\delta} =
    -2\pi\frac{1}{m}\frac{\Omega}{4\pi}N_\phi
    + (-\frac{1}{2} + \frac{1}{2m})\Omega \nonumber\\
  &\equiv& \gamma_{{\rm AB}}^{} + \gamma_{\rm Spin} \,.
\end{eqnarray}
{}From the last expression it is obvious that the quasihole has not
only a fractional charge $e^{*}=-e/m$, but also a total spin $S=
-\gamma_{\rm Spin}/\Omega = \textstyle{\frac{1}{2}}-1/2m$, where the
first term is due to kinematical effects, or equivalently to the shift
in the $N_e$--$N_\phi$ relation, and the second term is due to the
self-interaction.

For the quasielectron, the charge and the kinematical terms change
sign whereas the self-interaction term stays the same. (Note that the
statistics and self-interaction of the quasiholes and quasielectrons
have the {\em same\/} sign because they involve the product of their
charge and vorticity {\em both}\/ of which change sign between them.)
Consequently, we find,
\begin{equation} \label{eq:tot-berry-qe}
 \gamma_{{\rm B}}^{\rm qe} = 2\pi\frac{1}{m}\frac{\Omega}{4\pi}N_\phi
    +(\frac{1}{2} + \frac{1}{2m})\Omega \equiv \gamma_{{\rm AB}}^{} +
    \gamma_{\rm Spin} \Sp.
\end{equation}
To summarize: from (\ref{eq:tot-berry-qh}) and (\ref{eq:tot-berry-qe})
it is clear that the spin of the quasiparticles has a
charge-conjugation odd part that arises from the shift in the
number-flux relation and an even part, arising {}from the
self-interaction, that fulfills the generalized spin-statistics
relation (\ref{eq:gen-sp-stat}).

These results agree with the errata of Wen and Zee~\cite{Wen20}, and
with Lee and Wen~\cite{Lee9} and also with Li~\cite{Li3}. The latter
also attempted to directly compute the Berry phase for quasiparticles
on the sphere. Unfortunately, this last agreement is fortuitous
because Li incorrectly ignored the effect of the presence of the
quasiparticles on the background density {\em and}\/ ignored their
self-interaction---two errors that happen to cancel each other.

Finally, we digress to consider what happens when one puts another
(static) quasihole on the sphere. From the above considerations it
should be clear that there are three effects. The first two are an
increase of the total flux by one flux quantum and an increase in the
background density $\rho_0$ by $\frac{1}{4\pi m}$.  These effects
cancel each other in the sense that the background density expressed
in terms of $N_\phi$ is always $\rho_0 =
\frac{1}{4\pi}(\frac{1}{m}N_\phi + 1)$.  The third effect is a lack of
$1/m$ of an electron, either to the left or to the right of the
loop. If it is to the left, there will be a change in the enclosed
number of electrons and from (\ref{eq:encl-dens}) one has
$\Delta\gamma_{{\rm B}}^{} = 2\pi/m$, while if to the right,
$\Delta\gamma_{{\rm B}}^{} = 0$.  {}From the work by Hansson {\it et
al.\/}~\cite{Hanss2}, we know that the Berry phase $\Delta\gamma_{{\rm
B}}^{}$ is exactly opposite to twice the statistical phase $2\theta$,
and we thus have $\theta = -\textstyle{\frac{1}{2}}\Delta\gamma_{{\rm
B}}^{} = -\pi/m$. Hence, $\theta = S_{\rm even}/2\pi$ and the
generalized spin-statistics theorem (\ref{eq:gen-sp-stat}) is
satisfied, including signs.

\subsection{Berry-type calculation} \label{subsec:berrycalc}
A more straightforward way of calculating the total Berry phase for
the transport of a quasihole is to use the analogy with a spinor.  As
remarked earlier, the state with one quasihole (electron) at
$(\alpha,\beta)$ has a total angular momentum $L = N_e/2$, with a
maximal (minimal) projection in the direction given by
$(\alpha,\beta)$.  We can then use Berry's original calculation of the
geometric phase for a spin being turned around a closed loop
$C$~\cite{Berr1}.  If the spinor is always in an eigenstate with an
eigenvalue $S_\|$ along the direction of the magnetic field and the
direction of the field encloses a solid angle $\Omega$, the geometric
phase is given by Eq.~(\ref{eq:Berry-Berry}), $\gamma(\Omega) = -S_\|
\Omega$.  In our case, we have $S_\|=N_e/2$ for the quasihole, and
$S_\|=-N_e/2$ for the quasielectron, giving a total geometric phase
\FL
\begin{mathletters}
 \label{eq:berry}
\begin{eqnarray}
 \gamma_{{\rm B}}^{\rm qh} &=& -(\frac{N_e}{2})\Omega =
    -2\pi\frac{1}{m}\frac{\Omega}{4\pi}N_\phi + (-\frac{1}{2} +
    \frac{1}{2m})\Omega\Sp, \\ \gamma_{{\rm B}}^{\rm qe} &=&
    -(-\frac{N_e}{2})\Omega = 2\pi\frac{1}{m}\frac{\Omega}{4\pi}N_\phi
    + (\frac{1}{2} + \frac{1}{2m})\Omega\Sp,
\end{eqnarray}
\end{mathletters}
in agreement with our previous calculation. However, the present
derivation does not afford much insight into the physics of the
different terms. In any case, it should be reassuring for readers
worried about our numerical solution of the self-interaction integral
(\ref{eq:gamd}).

\subsection{Hierarchy states} \label{sec:hierarchy}

We can extend our calculation of the spin to states outside the
Laughlin sequence. For illustrative purposes we will use the technique
of the last section and restrict ourselves to the states in the
principal Jain sequences~\cite{Jain5}
\begin{equation} \label{eq:jainseq}
\nu = \frac{p}{2 n p + 1} \Sp.
\end{equation}
Following Jain, we view these states as consisting of $p$ filled
Landau levels of ``composite fermions'', each constructed from one
electron and $2 n$ flux quanta. For our purposes we need two
facts. First, the physical system with $N_e$ electrons and $N_\phi$
flux quanta is related to a composite fermion system with an equal
number ($N_e$) of composite fermions and the remaining $2 q^*$ flux
quanta by
\begin{equation} \label{eq:fluxreln}
N_\phi = 2 q^* + 2 n (N_e-1) \Sp;
\end{equation}
their wavefunctions are related by multiplication by the
Laughlin-Jastrow factor $ \prod_{ij} (z_i -z_j)^{2 n}$. Second, the
one-quasiparticle states of the physical system can be correctly
counted by counting the corresponding states of the composite fermion
system.

At $\nu = p/(2np +1)$, the $p$ Landau levels accommodate
\begin{equation} \label{eq:numparts}
N_e = \sum_{s=1}^{p} 2[q^*+(s-1)] +1 = 2pq^* + p^2
\end{equation}
composite fermions (electrons).  The one-quasihole states of the
physical system correspond to states with one composite fermion
removed from the $p$th Landau level.\footnote{We can also consider
vacancies in the lower Landau levels which correspond to vortices in
the lower condensates in the standard hierarchy construction. However,
these are {\em prima facie\/} higher energy excitations and, in fact,
do not form a well-defined band in numerical studies.} As there are
$2(q^*+p-1)+1$ of the latter states, it follows that the one-quasihole
states form a multiplet with $L = q^*+p-1$.

With these pieces of information, it is straightforward to write down
the Berry phase. Noting that Eq.~(\ref{eq:fluxreln}) holds for the
quasihole states, which have one fewer electron (i.e., $N_e = 2pq^*
+p^2 -1$), we find
\FL
\begin{equation} \label{eq:hierhole}
\gamma_{{\rm B}}^{\rm qh} = -L \Omega =
    -\frac{2\pi}{2np+1}\frac{\Omega}{4\pi}N_\phi - \left
    (\frac{2n-np^2}{2np+1} + p -1\right)\Omega\Sp.
\end{equation}
Similarly, quasielectrons are created by adding one composite fermion
to the $(p+1)$st Landau level and form an $L = q^*+p$ multiplet, which
leads to the Berry phase
\FL
\begin{equation} \label{eq:hierelec}
\gamma_{{\rm B}}^{\rm qe} = +L \Omega =
    +\frac{2\pi}{2np+1}\frac{\Omega}{4\pi}N_\phi + \left
    (\frac{-np^2}{2np+1} + p \right )\Omega\Sp.
\end{equation}
As a check, one can put $p=1$ and $2n+1=m$ to see that the above
results agree with the previous results for the Laughlin states.  In
the general case, we confirm from (\ref{eq:hierhole}) and
(\ref{eq:hierelec}) that the quasiparticles have a charge
$|e^*/e|=1/(2np+1)$, and we conclude that they have an intrinsic spin
\begin{equation}
S_{\rm even} = -\frac{1}{2} \frac{2n(p-1) +1}{2np+1} \Sp,
\end{equation}
in agreement with the generalized spin-statistics
relation.\footnote{This can be verified, for instance, from the
expression for the statistics at a general filling factor given by
Su~\cite{Su1}: at $\nu=r/s$, $-\theta/\pi = (n_1^2 r + n_2^2 s)/s
\pmod{2}$, where $n_1 r + n_2 s =1$. For the principal Jain states,
$r=p$, $s=2np+1$, whence $n_1=-2n$, $n_2=1$ and $\theta/\pi = -2n -
[2n(p-1) +1]/(2np+1) = 2 S_{\rm even}\pmod{2}$.}

\section{Discussion} \label{sec:disc}

Let us briefly recapitulate our analysis. We began by noting that
fractional-statistics particles may in general possess a fractional
spin, related to their statistics by the generalized spin-statistics
relation, that couples either to the self-linking of their world lines
or to the curvature of the surface they inhabit or both. For the
quasiparticles in the FQHE we observed that a coupling to the
curvature was essential for consistent quantization. We calculated the
Berry phase for a single quasiparticle on the sphere and found two
terms that couple to the curvature: a term odd under charge
conjugation (the orbital spin) that comes from the flux-number shift
and an even term (the intrinsic spin) that is due to the
self-interaction of the density profile of the quasiparticle and
agrees with the spin-statistics relation. On the plane, the orbital
spin vanishes trivially, while the self-interaction also vanishes
implying that there is no self-linking contribution to the
quasiparticle actions. Consequently, we believe, in disagreement
with~\cite{Lee9}, that the quasiparticle spin has no dynamical
consequences for planar systems.

In the foregoing we have assumed that we were dealing with spinless
electrons, i.e., we ignored their physical ${\rm SU(2)}$ spin.  For QH
systems whose ground states and excitations are fully polarized this
is perfectly legitimate as the corresponding wavefunctions factorize
into an orbital part and a trivial spin part (all spins up) which does
not contribute to any of the Berry phases. We remind the reader that
in studying QH systems on the sphere, it is standard practice to
translate the orbital effects of the transverse magnetic field on the
plane into the action of a monopole field while leaving the Zeeman
piece of the Hamiltonian intact. The alternative, coupling the
electron spins to the monopole field as well, introduces a coupling of
the intrinsic electron spin to the curvature which corresponds to a
spurious spin-orbit interaction in the planar problem.\footnote{As a
result the quasiparticle spins are modified but the generalized
spin-statistics relation still holds.  The monopole polarization can
effectively be described by reducing the total flux by
unity~\cite{Poly1}.  For the Haldane-Laughlin states, this gives
$N_\phi - 1 = m(N_e-1) + N_{\rm qh} - N_{\rm qe}$, and an additional
odd piece to the spin $\pm 1/(2m)$. This is the same result as one
would get if one na\"{\i}vely assumed that there was a deficiency
(excess) of $1/m$ electron spin.  Since this part is odd under charge
conjugation, it does not violate the generalized spin-statistics
relation.}

The simplest non-trivial case involves fully polarized ground states
but spin-reversed quasiparticles. Here it is not hard to see that the
reversed spins contribute to the odd part of the quasiparticle spin
without affecting the even part. In the framework of
Section~\ref{subsec:berrycalc}, we are interested in the magnitude of
$L$ for the quasiparticle states; as the location of the quasiparticle
is registered only by the expectation values of ${\bf L}$, and since
${\bf L}$ commutes with the electron spin $S_e$, the Berry phase for a
closed loop is still $-L \Omega$. As $L$ is decreased from its value
for a fully polarized quasiparticle by the number of reversed spins
$r$~\cite{Rezayi1}, the conclusion follows.  For more general cases,
where the ground state itself is partially polarized or unpolarized,
the spin enters in a more fundamental way into the dynamics, and the
analysis is more involved.  While we have not investigated these in
detail, we believe that our analysis in this paper generalizes
straightforwardly.

\section*{Acknowledgements}
We thank P. Elmfors, E. Fradkin, A. Goldhaber, T.~H. Hansson,
S. Kivelson, and M. Stone for valuable discussions.  This work was
supported in part by the Swedish Natural Science Research Council
(TE), by NSF grants Nos.\ DMR--9122385 and DMR--9157018 (SLS),
DMR--9416906 (SMG), and DMR--8957993 (DPA).

\appendix
\section*{The self-interaction integral}

In this appendix we derive the result of Eq.~(\ref{eq:gamd}) in the
limiting case of a small quasiparticle.  We have that the
self-interaction part of the Berry phase is
\FL
\begin{eqnarray}
\gamma_{{\rm B}}^{\delta} &=& \frac{1}{m(1-\cos\beta)}
\intl_{-\beta}^\beta\! d\alpha\,\sin(\theta_0+\alpha)\,
\intl_{-\varphi_1(\alpha)}^{\varphi_1(\alpha)} \!\!\!
\frac{d\varphi}{1-B(\alpha) e^{i\varphi}}\nonumber\\
&=&\frac{2}{m(1-\cos\beta)} \intl_0^\beta\!d\alpha\,\left(\bvph
\sin(\theta_0-\alpha)\times\right.\nonumber\\
&&\left\{\varphi_1(-\alpha)+ \arctan \left[
\frac{\sin\varphi_1(-\alpha)} {B^{-1}(-\alpha)-\cos\varphi_1(-\alpha)}
\right] \right\}\nonumber\\
&&-\sin(\theta_0+\alpha)\,\arctan\left[\frac{\sin\varphi_1(\alpha)}
{B(\alpha)-\cos\varphi_1(\alpha)}\right]\left.\bvph\right) \Sp,
\label{eq:app1}
\end{eqnarray}
where
\begin{equation}
B(\alpha)=\frac{1+\cot{\frac{\theta_0}{2}}\tan{\frac{\alpha}{2}}}
{1-\cot\frac{\theta_0}{2}\tan{\frac{\alpha}{2}}} \Sp,
\end{equation}
and
\begin{equation}
\cos\varphi_1=\frac{\cos\beta-\cos\theta_0\cos(\theta_0+\alpha)}
{\sin\theta_0\sin(\theta_0+\alpha)}\Sp.
\end{equation}
(It is important to separately consider the $-\beta\leq\alpha\leq 0$
and $0\leq\alpha\leq\beta$ portions of the integral in
Eq.~(\ref{eq:app1}), owing to the fact that $|B(\alpha)|<1$ in the
former region, while $|B(\alpha)|>1$ in the latter region.)

In the limit $\beta\ll 1$, we can expand to obtain
\begin{equation}
\varphi_1(\alpha)=\sqrt{\beta^2-\alpha^2}\csc(\theta_0)\,
[1-\textstyle{\frac{1}{2}}\alpha\cot(\theta_0)] +{\cal
O}(\alpha^4,\beta^4) \Sp,
\end{equation}
and
\begin{mathletters}
\begin{eqnarray}
B(\alpha)&=&1+\alpha\csc\theta_0+\textstyle{\frac{1}{4}}\alpha^2\sec^2
\textstyle{\frac{\theta_0}{2}}+ {\cal O}(\alpha^3)\Sp, \\
B^{-1}(-\alpha)&=&1+\alpha\csc\theta_0
+\textstyle{\frac{1}{4}}\alpha^2\csc^2
\textstyle{\frac{\theta_0}{2}} + {\cal O}(\alpha^3)\Sp.
\end{eqnarray}
\end{mathletters}
Note that since $\alpha$ runs from $-\beta$ to $\beta$, an expansion
in powers of $\alpha$ is tantamount to an expansion in powers of
$\beta$.  Thus, for our purposes, $\alpha$ and $\beta$ may be regarded
as of the same degree of smallness.  {}From these results, we derive
\begin{eqnarray}
\frac{\sin\varphi_1(\alpha)}{B(\alpha)-\cos\varphi_1(\alpha)}&=&\svph
\frac{\sin\varphi_1(-\alpha)}{B^{-1}(-\alpha)-\cos\varphi_1(-\alpha)} \\
&=&{\sqrt{\beta^2-\alpha^2}\over\alpha}\left [1-{\beta^2\over 2\alpha}
\csc\theta_0+\ldots\right ]\Sp,\nonumber
\end{eqnarray}
which, when inserted into Eq.~(\ref{eq:app1}) together with an
evaluation to first order in $\alpha$ and $\beta$ and the change of
variables $\alpha=\beta\sin(\omega)$, leave us with elementary
integrals yielding $\gamma_{{\rm B}}^{\delta}{}=\Omega/2m$, as claimed
in Eq.~(\ref{eq:gamS}).

\end{document}